\documentclass[floatfix,notitlepage,twocolumn,longbibliography]{revtex4-1}%
\usepackage{graphicx}%
\usepackage{amsmath}%
\usepackage{amsfonts}%
\usepackage{amssymb}
\usepackage{bm}
\usepackage{comment}
\usepackage{color}
\usepackage[breaklinks]{hyperref}
\usepackage[normalem]{ulem}

\def\d{{\partial}}
\def\s{{\sigma}}

\def\k{{ {\bm k} }}

\def\0{{ {\bm 0} }}

\def\ve{{\varepsilon}}

\def\rR{{ {\rm R} }}
\def\rA{{ {\rm A} }}

\def\cV{{ {\cal V} }}

\def\ang{{ \mathrm{\mathring{A}} }}

\usepackage{xcolor}

\allowdisplaybreaks[4]

\begin{document} 
\title{Nonlinear charge transport properties in chiral tellurium}
\author{ 
Kazuki Nakazawa$^1$,  
Terufumi Yamaguchi$^1$, and 
Ai yamakage$^2$ 
} 
\address{
$^1$RIKEN Center for Emergent Matter Science (CEMS), 2-1 Hirosawa, Wako, Saitama 351-0198, Japan
\\
$^2$Department of Physics, Nagoya University, Nagoya 464-8602, Japan
}

\date{\today}

\begin{abstract}
Extensive research has focused on phenomena arising from the chirality of crystalline or magnetic structures. Recently, we have proposed the \lq\lq Nonlinear Chiral Thermo-Electric (NCTE) Hall effect," in which current flows in the direction of the cross product of the electric field and the temperature gradient in a material with a chiral structure. Despite its importance for device applications such as heat flow sensors and logic circuits, there are no quantitative demonstrations in actual materials. We investigate the NCTE Hall effect and the second-order response to the electric field in Te, which hosts the chiral crystal structure, based on {\it ab initio} calculations and symmetry argument. We clarify that the NCTE Hall current can flow parallel to the chiral axis, whereas such behavior is absent for the second-order response to the DC electric field. Quantitative estimation predicts the large NCTE Hall current which is sufficiently detectable in experiments, and the orbital magnetic moment will play an important role in the vicinity of the top of the valence band. 
\end{abstract}

\maketitle

Chirality is one of the fundamental degrees of freedom in nature~\cite{Greed2022,Ozturk2022}. In particular, the chirality of crystals and magnetic structures in condensed matter has attracted much attention because of the novel phenomena they produce. For example, magnetochiral dichroism~\cite{Rikken1998}, chirality induced spin selectivity (CISS)~\cite{Ray1999,Naaman2019}, and chiral phonons~\cite{Ishito2023,Ueda2023,Ohe2024} are representative examples of phenomena caused by the chirality of crystals, which are currently under intensive study. 

Besides, the second-order response to an external field is linked to the breaking of spatial inversion symmetry that resides in the materials. For instance, both time-reversal and inversion symmetry breaking give rise to the band asymmetry leading to the second-order response to the electric field~\cite{Rikken2001,Yokouchi2017,Ishizuka2020,DLBLK}, and the exhibition of the Berry curvature dipole due to the inversion symmetry breaking also leads to the second-order response~\cite{SF2015,Kang2019,Ma2019,DLX2021,DWSLX}. The inclusive understanding of such nonlinear transport is summarized in terms of the band geometry~\cite{GYN2014,MN}, governed from the microscopic analyses~\cite{MP}. 

Meanwhile, the \lq\lq Nonlinear Chiral Thermo-Electric (NCTE) Hall effect," in which current flows in the direction of the cross product of the electric field and the temperature gradient, have been studied theoretically~\cite{Hidaka2018,Nakai2019,Toshio2020,YNY2023}, which is unique to the chiral materials without any mirror or inversion symmetry. 
The microscopic calculation revealed that the NCTE Hall current $\langle j_z^{\rm NCTE} \rangle = \sigma_z^{\rm NCTE} \{ {\bm E} \times (- \nabla T/T ) \}_z$ is dominated by following two terms~\cite{YNY2023};  
\begin{align} 
&\sigma_z^{\rm NCTE} \simeq \sigma^{\mathrm{BC}}_{z} + \sigma^{\mathrm{OM}}_{z} \label{eq:NCTE_TOT}, \\
&\sigma^{\mathrm{BC}}_{z}= e^{2} \tau \sum_{n,\bm{k}}
\left( \varepsilon_{n\bm{k}} - \mu \right) \left( - \frac{\partial f}{\partial \varepsilon} \right)_{\varepsilon = \varepsilon_{n\bm{k}}}
\notag \\
&\quad \quad \quad \times
\left[ 	v_{n\bm{k}}^z \Omega^{z}_{n\k} - \frac{1}{2} \left\{ 
		v_{n\bm{k}}^x \Omega^{x}_{n\k} +
		v_{n\bm{k}}^y \Omega^{y}_{n\k} \right\} \right],
\label{eq:NCTE_BC}
\\
&\sigma^{\mathrm{OM}}_{z} = -\frac{1}{2} e \tau \sum_{n,\bm{k}}
\left( \varepsilon_{n\bm{k}} - \mu \right) \left( - \frac{\partial f}{\partial \varepsilon} \right)_{\varepsilon = \varepsilon_{n\bm{k}}} \bm{\nabla}_{\bm{k}} \cdot \bm{m}^{\perp}_{n\bm{k}} ,
\label{eq:NCTE_OM}
\end{align}
where ${\bm E}$ is the DC electric field, $T$ is the temperature, $e$, $\tau$, and $\mu$ are the elementary charge, electron lifetime, and chemical potential, respectively, $f$ is the Fermi-Dirac distribution function, and ${\bm \Omega}_{n \k}$ and ${\bm m}_{n \k}$ are the Berry curvature and the orbital magnetic moment, 
\begin{align}
{\bm \Omega}_{n\k} &= i \sum_{m \neq n} \frac{ \langle n \k | \hat{\bm \cV}_\k | m \k \rangle \times \langle m \k | \hat{\bm \cV}_\k | n \k \rangle }{ (\ve_{n\k} - \ve_{m\k} )^2 } , \label{eq:BC}
\\
{\bm m}_{n\k} &= \frac{ie}{2} \sum_{m \neq n} \frac{ \langle n\k | \hat{\bm \cV}_\k | m \k \rangle \times \langle m \k |  \hat{\bm \cV}_\k | n \k \rangle }{ (\ve_{n\k} - \ve_{m\k} ) } , \label{eq:OM}
\end{align} 
respectively, with an eigenenergy $\ve_{\k n}$ and an eigenvector $| n\k \rangle$ of the Hamiltonian $\hat{H}_\k$, the velocity operator $\hat{\bm \cV}_\k = \nabla_\k \hat{H}_\k$, and group velocity ${\bm v}_{n\k} = \nabla_\k \ve_{n\k}$.
\begin{figure}
\includegraphics[width=85mm]{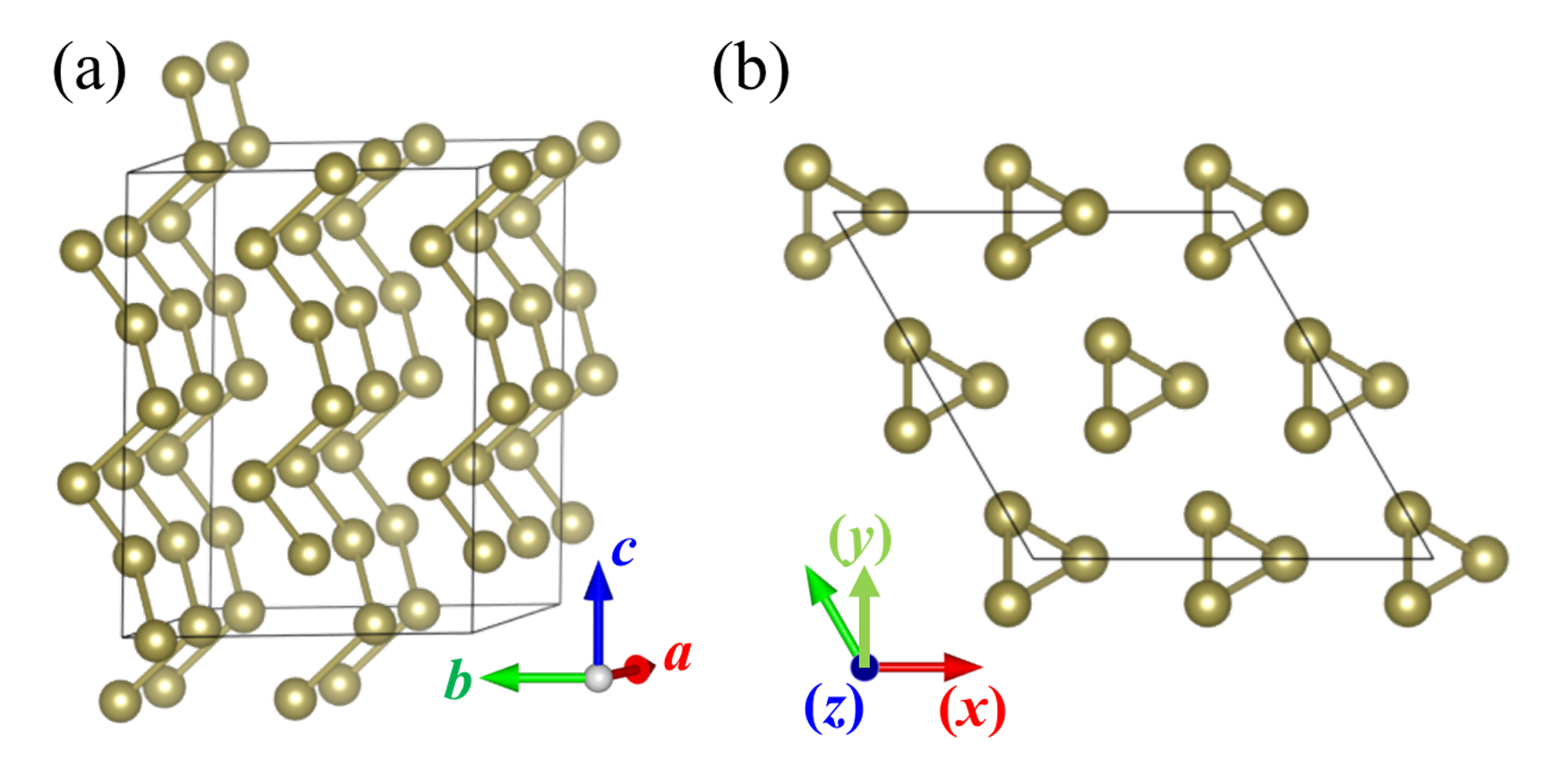}
\caption{(a) Lattice structure of the tellurium. Black lines represent the extended unit cell which contains eight primitive unit cells. (b) Top view of (a). We denoted the definition of $x$, $y$, and $z$ axes. The crystal structures are visualized by VESTA~\cite{Momma2011}.
}
\label{fig:1}
\end{figure}
\begin{figure*}
\includegraphics[width=176mm]{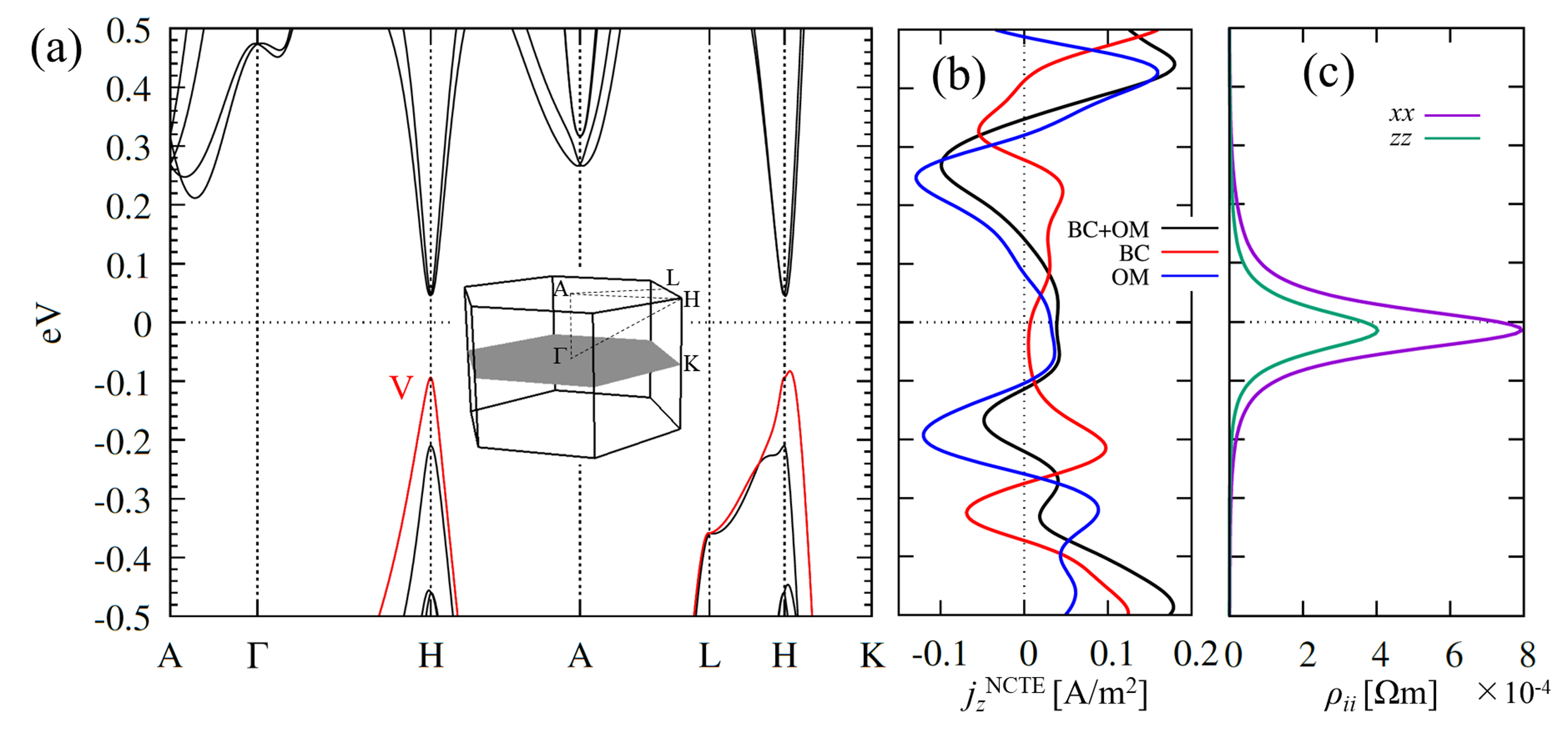}
\vspace{-3mm}
\caption{(a) Band structure of the tellurium obtained from the DFT calculation. The red line specifies the highest-energy valence band V, which we focus on. The inset shows the Brillouin zone, where the gray hexagon represents the $k_z = 0$ plane. (b) Chemical potential dependence of NCTE Hall current density. Black, red, and blue lines represent total, Berry curvature (BC), and orbital magnetic moment (OM) contributions, respectively. (c) Longitudinal resistivity with respect to $x$ and $z$ directions.}
\label{fig:2}
\end{figure*}
We here set the Dirac (Planck) constant as $\hbar = 1 \ (h = (2\pi)^{-1})$. This NCTE Hall effect inverts its sign depending on the chirality~\cite{YNY2023}, hence it can detect the handedness of the lattice/magnetic structure. Moreover, it also has the potential for device applications, such as heat flow sensors or logic circuits. However, theoretical and experimental demonstrations considering actual materials are yet to be proposed, and hence, quantitative evaluation and characterization of the NCTE Hall current will be important. 

The tellurium (Te) crystal belongs to the space group $P3_1 21$ (No.~152) or $P3_2 21$ (No.~154) which contains a three-fold chiral axis and does not have any mirror or spatial inversion symmetry [Fig.~\ref{fig:1}]. While Te has various applications as a compound, the physical properties of Te itself have recently attracted attention such as a thermoelectric conversion due to the semiconducting property~\cite{Lin2016}, characteristic structure of the Berry curvature related to the existence of Weyl points~\cite{Hirayama2015}, and nonreciprocal transport phenomena under magnetic fields due to its crystal chirality~\cite{Hirobe2022,Calavalle2022}. Moreover, recently, the orbital moment and its consequences have attracted attention~\cite{Yoda2015,Yoda2018}. In particular, the effective model analysis predicts the hedgehog-like structure of the orbital moment at the top of the valence band~\cite{Maruggi2023}, which motivates us to study whether the orbital magnetic moment plays a role in the NCTE Hall effect or not.

In this paper, we investigate the NCTE Hall effect and the second-order response to the electric field in tellurium based on {\it ab initio} calculations. We clarify that the NCTE Hall current is finite while the second-order response to the DC electric field is absent along the chiral axis. Our quantitative estimation predicts that the NCTE Hall current is large enough to observe experimentally, and the role of the Berry curvature and orbital magnetic moment is clarified. 

The OpenMX code~\cite{OpenMX,Ozaki2003} is used to obtain the band structures based on the density functional theory (DFT). The generalized gradient approximation proposed by Perdew-Burke-Ernzerhof~\cite{PBE1996} is employed for the exchange correlation functional. The norm-conserving and total angular momentum-dependent pseudopotentials are employed, and the pseudo-atomic orbital Te7.0-s4p3d3f2 is chosen as a basis of the Kohn-Sham wave functions. We perform the fully relativistic calculation to incorporate the effect of spin-orbit coupling. We used the right-handed crystal of the space group $P3_1 21$, and set the lattice constant to $a=4.6~\ang$ and $c=5.9~\ang$ to reproduce the band gap; we will touch this point later. We set the cutoff energy which specifies the fast Fourier transform grid to 1200 Ry and sampled the Brillouin zone with $16^3$ $k$ points. 

The band structure in the window [$-0.5$~eV, 0.5~eV] is shown in Fig.~\ref{fig:2}(a). The band gap is present around the H point, where the lowest-energy conduction band and highest-energy valence band exhibit exotic characteristics such as Berry curvature (dipole) and orbital magnetic moments, as we will discuss later. The overall band structure reproduces well that of previous study~\cite{Hirayama2015}. Note that Ref~\cite{Hirayama2015} takes the experimental in-plane (perpendicular to chiral axis) lattice constant of $a \sim 4.456~\ang$~\cite{Keller1977,Adenis1989}, but considers the LDA+GW framework to implement the effect of electron-electron interaction to obtain the band gap. On the other hand, if we take the same $a$ within the GGA calculation, the band gap is closed and cannot reproduce the semiconducting property. Instead, we just take the larger in-plane lattice constant to reproduce the band gap. Actual Te samples are slightly doped with holes \cite{Lin2016}, leading to an estimated chemical potential of approximately $\mu \sim -0.1 \ \mathrm{eV}$, which corresponds almost to the top of the valence band. Henceforth we will focus on this value frequently.

\begin{figure}
\includegraphics[width=84mm]{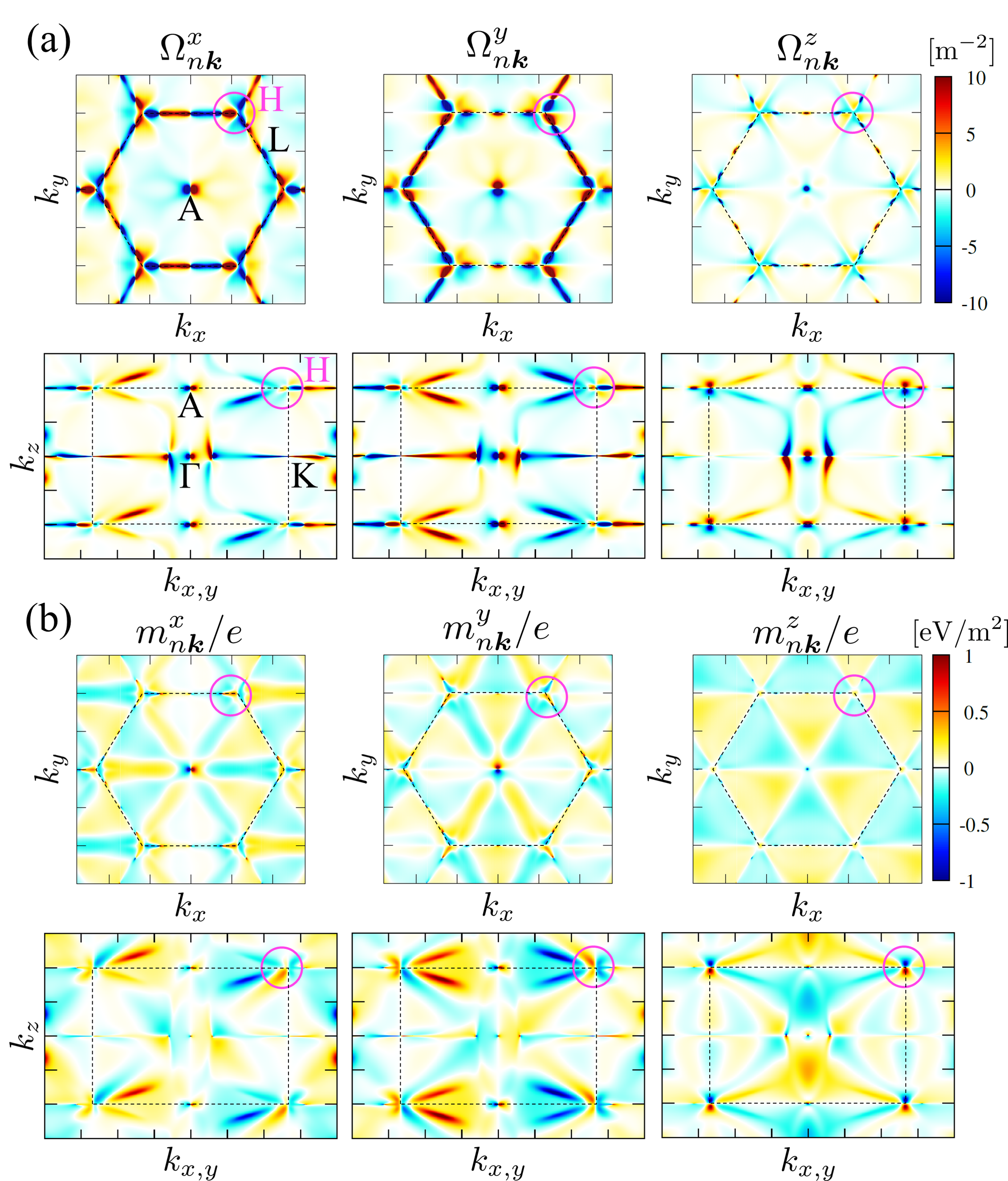}
\caption{
Momentum space distributions of (a) Berry curvature ${\bm \Omega}_{n\k}$ and (b) orbital angular momentum ${\bm m}_{n\k}$ of the band V, which are calculated using Eqs.~\eqref{eq:BC} and \eqref{eq:OM}, respectively. The dotted lines are projected boundaries of the Brillouin zone, with the high-symmetry points H, L, A, $\Gamma$, K, shown in the inset of Fig.~\ref{fig:2}(a). The horizontal axes in the second row of each panel represent $k_{x,y} = \sqrt{3} k_x = k_y$. The pink circles represent the region around the H point, which we mainly focus on in the main text. 
}
\label{fig:3}
\end{figure}

\begin{figure*}
\includegraphics[width=175mm]{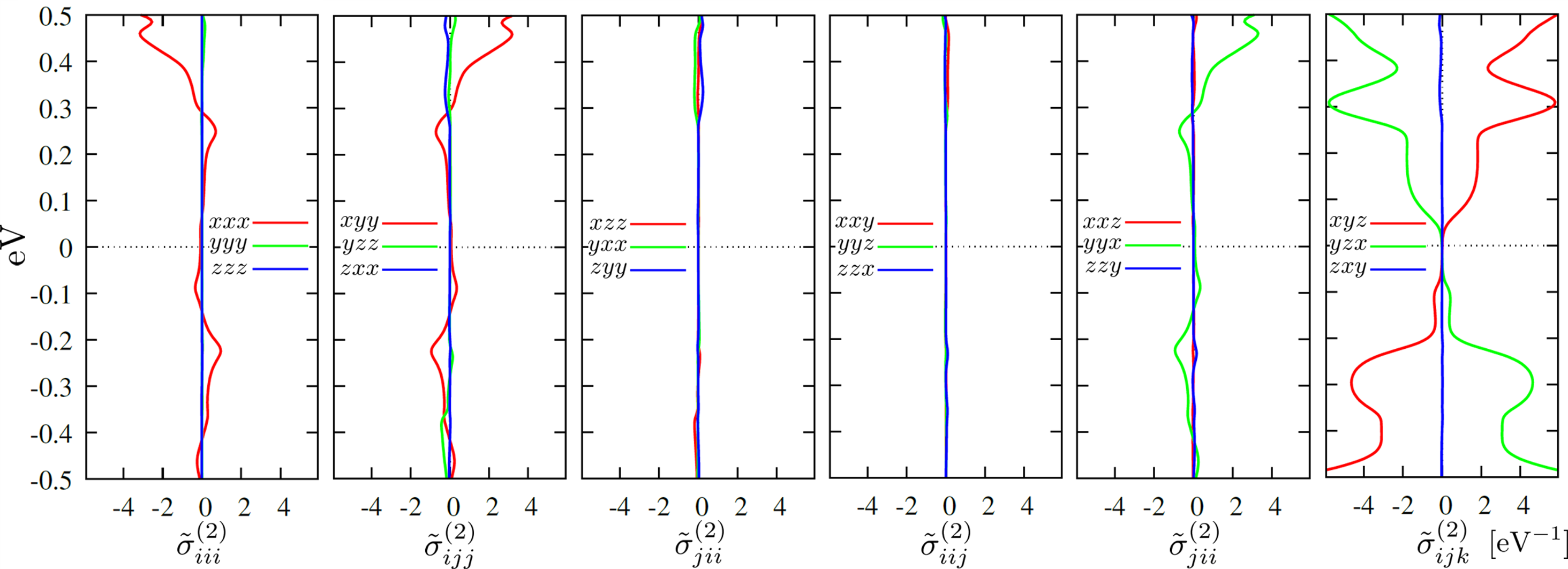}
\caption{Chemical potential dependence of the normalized second-order nonlinear conductivity $\tilde{\sigma}_{ijk}^{(2)} \equiv \sigma_{ijk}^{(2)}/(e^3/h)$. The red, green, and blue lines correspond to the current along the $x$, $y$, and $z$ directions, respectively. The small fluctuations seen in the components except for $\tilde{\sigma}_{xxx}^{(2)}$, $\tilde{\sigma}_{xyy}^{(2)}$, $\tilde{\sigma}_{yyx}^{(2)}$, $\tilde{\sigma}_{xyz}^{(2)}$, and $\tilde{\sigma}_{yzx}^{(2)}$ come from symmetry breaking due to the Wannierization. 
}
\label{fig:4}
\end{figure*}

To obtain the tight-binding model, we constructed maximally localized Wannier functions~\cite{MV1997,SMV2001} using OpenMX~\cite{WOT2009}. The eigenstates within the energy window of [$-6$~eV, 3.5~eV] are employed and projected to the $p$ orbitals for each atom, and hence in total, 18 orbitals are considered, including spin degrees of freedom. The Wannier model reproduces the original band structure obtained by DFT calculations.

The NCTE Hall current $j_z^{\rm NCTE}$ is calculated employing the Wannier model. We use Eqs.~\eqref{eq:NCTE_TOT}-\eqref{eq:NCTE_OM} to calculate NCTE Hall current defining the $x$, $y$, and $z$ axis parallel to ${\bm a}$, ${\bm c} \times {\bm a}$, and ${\bm c}$ directions, respectively [see Fig.~\ref{fig:1}(b)]. We take the $k$-mesh number of $240^3$ for the momentum integral. The parameters are set as $E_x = 1100$~V/m, $k_{\rm B}T = 0.03$~eV, $\d_y T/ T = 100~{\rm m}^{-1}$, $\tau = 10$~{\rm fs}. The chemical potential dependence of $j_z^{\rm NCTE}$ is shown in Fig.~\ref{fig:2}(b), representing the finite NCTE Hall effect. Comparing the value of $\sigma_z^{\rm BC}$ and $\sigma_z^{\rm OM}$, they overall show a comparable value within the energy range of $-0.5$ to  0.5~eV. Importantly, if we focus around the top of the valence band (at the vicinity of $\mu = -0.1$~eV), $\sigma_z^{\rm OM}$ is dominant while $\sigma_z^{\rm BC}$ is suppressed. This indicates that the emergence of the characteristic structure of orbital magnetic moment and the cancellation of the total Berry curvature dipole occurs around $\mu \sim -0.1$~eV simultaneously. We will later discuss these points with the actual distribution of ${\bm \Omega}_{n\k}$ and ${\bm m}_{n\k}$. Moreover, the obtained NCTE current density is valued around $0.04$~$\rm A/m^2$ at $\mu \sim -0.1$~eV, which is already measurable. To consider the possible range of the NCTE current, we calculate longitudinal resistivity $\rho_{ii}$ with respect to $x$ and $z$ directions [Fig.~\ref{fig:2}(c)]. Under $k_{\mathrm{B}}T=0.03~{\rm eV}$ and $\tau = 10$~fs, the resistivity is valued around $\sim 10^{-4}~\Omega {\rm m}$, which is comparable to the experimental value. Depending on the sample quality, the typical value of the resistivity is ranged between $10^{-3}$ to $10^{-6}~\Omega {\rm m}$~\cite{Ross1992,Abad2015,Lin2016,Rabadanov2021}. As electrical conductivity $\sigma_{ii} = 1/\rho_{ii}$ and NCTE current both scale as linear of $\tau$, the NCTE Hall current can be 10 to 100 times larger in the more conductive sample. 

\begin{table}[t]
\centering
\newlength{\height} 
\setlength{\height}{3mm}
\begin{center}
\fontsize{10pt}{13pt}\selectfont
\begin{tabular}{ccccll}
 \hline\hline
 $D_3$ & $E$ & $2C_3$ & $3C_2'$  & Linear & Quadratic
\\
\hline
 $A_1$ & 1 & 1 & 1 &  & $XX+YY$, $ZZ$
\\
 $A_2$ & 1 & 1 & $-1$ & $Z$ & $XY-YX$
 \\
 $E$ & 2 & $-1$ & 0 & $(X,Y)$ & $(XX-YY, -XY-YX)$, 
\\
& & & & &  $(YZ,-XZ)$, $(ZY,-ZX)$
\\
\hline\hline
\end{tabular}
\caption{Character table for the point group $D_3$. $A_1$, $A_2$, and $E$ represent irreducible representations, and $X$, $Y$, and $Z$ are linear and quadratic basis functions corresponding to the coordinate $x$, $y$, and $z$. }
\label{tab:D3}
\end{center}
\end{table}

Figure~\ref{fig:3} shows the momentum space distributions of the Berry curvature ${\bm \Omega}_{n\k}$ and the orbital magnetic moment ${\bm m}_{n\k}$ of the band V. Figure~\ref{fig:3}(a) is plots of each component of ${\bm \Omega}_{n\k}$. We observe the dipole-like properties in each component around the H point. However, due to its isotropy of the momentum space distribution, Berry curvature (dipole) satisfies the relation ${\sum_\k v_{n \boldsymbol{k}}^x \Omega_{n \boldsymbol{k}}^x \sim \sum_\k v_{n \boldsymbol{k}}^y \Omega_{n \boldsymbol{k}}^y \sim \sum_\k v_{n \boldsymbol{k}}^z \Omega_{n \boldsymbol{k}}^z}$, leading to the suppression of $\sigma_z^{\rm BC}$ around $\mu \sim -0.1$~eV. On the other hand, ${\bm m}_{n\k}$ shown in Fig.~\ref{fig:3}(b) shows a similar characteristic mapping around the H point. However, the NCTE Hall effect only captures the properties of the in-plane component, namely ${\bm m}_{n\k}^\perp = (m_{n\k}^x , m_{n\k}^y , 0)$ with the in-plane texture such that $\nabla_\k \cdot {\bm m}_{n \k}^\perp \neq 0$ which prevent the cancellation. This situation is quite similar to that in the minimal model discussed in Ref.~\citenum{YNY2023}. Generally, $\sigma_z^{\rm BC}$ should be suppressed and only $\sigma_z^{\rm OM}$ can be only visible in the isotropic and/or the cubic systems.  

For the comparison, the second-order current response to the DC electric field $\langle j_i^{(2)} \rangle = \sigma_{ijk}^{(2)} E_j E_k$ are calculated in the same Wannier model. We here consider the following expression of the second-order DC nonlinear conductivity $\sigma_{ijk}^{(2)}$~\cite{MP,MN,note1}; 
\begin{align}
&\sigma_{ijk}^{(2)} \simeq \frac{2e^3}{V} \int \frac{d\ve}{2\pi} \left(-\frac{\partial f}{\partial \ve}\right)  
\nonumber \\
&\quad \times {\rm Im} \sum_\k {\rm tr} \Biggl\{ \hat{\cV}_i \frac{\partial \hat{G}^\rR }{\partial \ve} \left( \hat{\cV}_j \hat{G}^\rR \hat{\cV}_k + \frac{1}{2} \hat{\cV}_{jk} \right) ( \hat{G}^\rR - \hat{G}^\rA ) \Biggr\}  
\nonumber \\
&\quad + (j \leftrightarrow k), 
\label{eq:NLC}
\end{align}
where $\hat{G}^\rR = (\ve - \hat{H}_\k - \hat{\Sigma}^\rR)^{-1} = (\hat{G}^\rA)^\dagger$ is retarded Green's function with self-energy $\hat{\Sigma}^{{\rm R}}$, and $\hat{\cV}_{ij} = \partial_{k_i} \partial_{k_j} \hat{H}_\k$. The trace runs over all of the orbital/band indices. For simplicity, we here consider the constant pure imaginary self-energy $\hat{\Sigma}^{\rm R}= -i/(2\tau)$. The chemical potential dependence of normalized second-order nonlinear conductivity $\tilde{\sigma}_{ijk}^{(2)} = \sigma^{(2)}_{ijk}/(e^3/h)$ is shown in Fig.~\ref{fig:4}. We observe only 5 finite components out of 18 components and found approximate relations $\sigma_{xxx}^{(2)} = -\sigma_{xyy}^{(2)} = -\sigma_{yyx}^{(2)}$ and $\sigma_{xyz}^{(2)} = -\sigma_{yzx}^{(2)}$; there are only two independent tensor components. Moreover, strikingly, the nonlinear response along the $z$ direction (chiral axis) never arises, $\sigma_{zjk}^{(2)}=0$, if we only apply DC electric fields, showing the peculiarity of the NCTE Hall current. 

To verify the results of the NCTE Hall effect and second-order DC nonlinear conductivity, we perform the symmetry analysis in the case of $D_3$ point group in which the tellurium crystal belongs, summarized in Table~\ref{tab:D3}. Due to the symmetry, in the equation $J_i = \sigma^{(2)}_{ijk} E_j E_k$, the left and right sides are represented by the same basis of irreducible representation. As a result, the possible combinations of applied fields and currents are
\begin{align}
(J_x, J_y, J_z) &= \sigma_1 (E_x^2-E_y^2, -2 E_x E_y, 0) \nonumber \\ 
 &+ \sigma_2 (E_yE_z, -E_x E_z, 0),
\end{align}
for the second-order response to the electric field, and
\begin{align}
(J_x, J_y, J_z) &=   
\sigma'_1 (0, 0, E_x \partial_y T - E_y \partial_x T) \nonumber \\
&+ \sigma'_2 (E_x \partial_x T - E_y \partial_y T, -E_x \partial_y T - E_y \partial_x T, 0) \nonumber \\ 
&+ \s'_3 (E_y \partial_z T, -E_x \partial_z T, 0) \nonumber \\
&+ \s'_4 (E_z \partial_y T, -E_z \partial_x T, 0)
\end{align} 
for the response to the product of the electric field and the temperature gradient. Here $\s_i$ and $\s'_i$ are the corresponding coefficients. $J_z$ only appears in $\s'_1 = \s_z^{\rm NCTE}$, as a response to the cross product of the electric field and temperature gradient applied to the $xy$-plane, representing the NCTE Hall effect. The number of the independent tensor components is two for the response to $E_i E_j$ as we already see in the {\it ab initio} results, while that is four for the $E_i \partial_j T$ implying the variety of the response and motivating the analysis of the general response to the product of the electric field and the temperature gradient including symmetric components other than the NCTE Hall effect. A more generalized implication from the symmetry argument is that the second-order electric current response along the chiral axis requires two kinds of applied field; for example, we could employ a chemical potential gradient instead of the temperature gradient. Moreover, if we extend the consideration to the AC applied field, the circularly polarized light also induces the current as $J_z \propto \boldsymbol{E} \times \boldsymbol{E}^*$ for ${\bm E} = (1,i,0)$~\cite{Ma2022}. 

To summarize, we investigated the NCTE Hall effect, the flow of the current to the direction of the cross product of the electric field and the temperature gradient, in the tellurium utilizing the {\it ab initio} calculation and symmetry arguments. The Berry curvature dipole and orbital magnetic moment will overall equally contribute to the NCTE Hall effect. Especially, the orbital magnetic moment is dominant around the band top of the valence band, while Berry curvature contribution is canceled because of the isotropy around the H point. The quantitative estimate of the NCTE Hall current density predicts that its amount is sufficiently measurable in the experiment. Comparison to the second-order response to the DC electric field revealed that the current along the chiral axis is only expected for the NCTE Hall effect but does not occur for the second-order nonlinear conductivity. This is verified by the symmetry analysis, which predicts the diversity of the setups, e.g., applying chemical potential gradient as well as circularly polarized light to obtain the nonlinear response along the chiral axis.  

The authors thank Y. Araki, H. Chudo, F. Kagawa, T. Kikkawa, J. Ieda, M. Imai, T. Nomoto, Y. Michishita, T. Morimoto, T. Sato, and M. Umeda for fruitful discussions. Parts of the numerical calculations have been done using the Supercomputer HOKUSAI BigWaterfall2 (HBW2), RIKEN. This work is supported by JSPS KAKENHI (Grant Nos.~JP20K03835, JP21K14526, and JP21K13875).

\end{document}